\documentclass[journal=nalefd,manuscript=article,layout=traditional]{achemso}    

\usepackage{braket}
\usepackage[version=3]{mhchem}
\usepackage{graphicx}
\usepackage{upgreek} 
\setkeys{acs}{maxauthors=999}

\author{Gbenga S. Agunbiade}
\affiliation{Department of Physics and Astronomy, The University of Kansas, Lawrence, Kansas 66045, United States}

\author{Ting Zheng}
\affiliation{Department of Physics and Astronomy, The University of Kansas, Lawrence, Kansas 66045, United States}

\author{Hui Zhao}
\affiliation{Department of Physics and Astronomy, The University of Kansas, Lawrence, Kansas 66045, United States}
\email{huizhao@ku.edu}

\title[\texttt{achemso} [Stacking Polarity–Controlled Interlayer Photocarrier Dynamics in MoSe$_2$/MoS$_2$ Heterostructures]
{Stacking Polarity–Controlled Interlayer Photocarrier Dynamics in MoSe$_2$/MoS$_2$ Heterostructures}


\begin{document}

\begin{tocentry}
  \includegraphics[width=8.5cm]{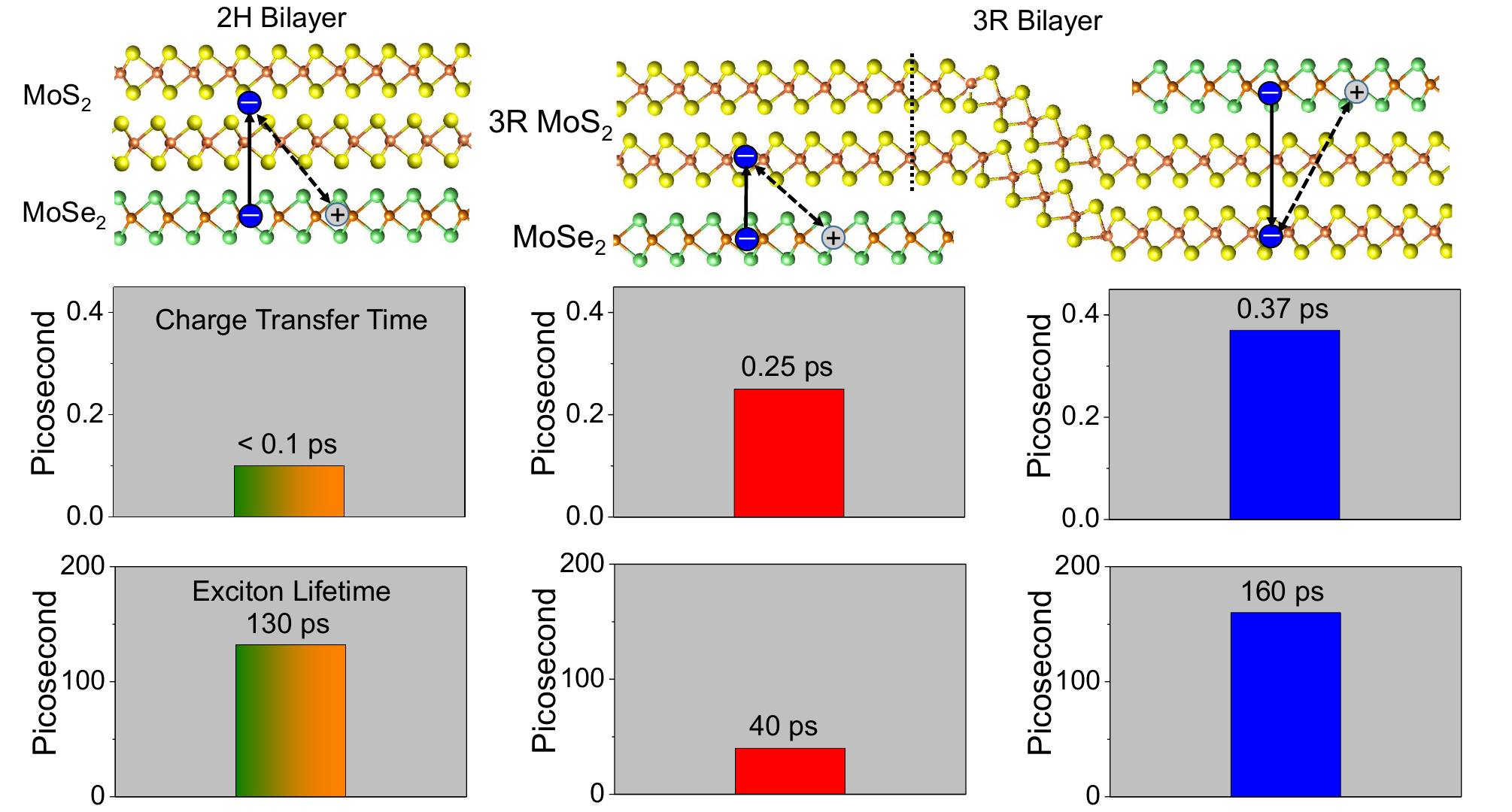}
\end{tocentry}

\begin{abstract}

Control of interlayer photocarrier dynamics is central to optoelectronic applications of van der Waals heterostructures, yet deterministic and spatially uniform tuning strategies remain limited. Here we show that stacking polarity provides a global control parameter for photocarrier dynamics in MoSe$_2$/MoS$_2$ heterostructures. By comparing hexagonal (2H) and rhombohedral (3R) MoS$_2$ bilayers and engineering opposite interface terminations in 3R stacking, we resolve stacking-dependent interlayer charge-transfer dynamics using ultrafast pump--probe spectroscopy. While charge transfer in the 2H heterostructure occurs faster than the experimental resolution, the 3R heterostructures show time-resolvable charge transfer that slows from $0.25$ to $0.37$~ps depending on stacking polarity. Furthermore, the interlayer exciton lifetime is tuned from $\sim$$40$ to $\sim$$170$~ps. These effects arise from stacking-induced layer polarization in 3R MoS$_2$, which modulates interfacial wavefunction overlap.

\end{abstract}

The ability to control photocarrier generation, transport, and recombination dynamics is central to optoelectronic technologies, including photodetectors, photovoltaics, and light-emitting devices. In conventional semiconductors, these processes are largely intrinsic material properties dictated by chemical composition and crystal symmetry, offering limited post-synthesis tunability. Chemical alloying and heteroatomic substitution provide routes to modify electronic structure and carrier lifetimes; however, such approaches inevitably alter multiple material parameters simultaneously and may introduce disorder or inhomogeneity. 

More recently, van der Waals heterostructures based on two-dimensional (2D) materials have opened new opportunities for engineering charge transport and recombination.\cite{nn7699,2dm3042001,n499419,nrm116042,s353461} In particular, moir\'e superlattices in twisted heterobilayers have demonstrated spatially modulated interlayer coupling and emergent correlated phenomena.\cite{n56766,n56771,n56776} However, moir\'e engineering intrinsically produces in-plane variations of electronic properties and requires precise angular control, which may limit scalability and uniform device implementation. A critical need therefore remains for deterministic and globally uniform strategies to tailor interlayer photocarrier dynamics without introducing chemical disorder or spatial inhomogeneity.

Layered transition metal dichalcogenides (TMDs) provide a versatile platform for stacking-engineered functionality because multiple polytypes with distinct stacking sequences can exist within the same chemical composition. While most previous studies have focused on the hexagonal (2H) phase, rhombohedral (3R) TMDs have recently attracted attention due to their broken inversion symmetry and stacking-induced electronic asymmetry.\cite{nc112391,am301704674,x12041005} Unlike 2H bilayers, which preserve inversion symmetry and yield layer-symmetric electronic states, 3R stacking generates inequivalent layers and can produce layer-polarized conduction and valence band states.\cite{nc112391,b89075409,nn9611,np16469,sa8eade3759,nn1836,b109035410,acsnano1941244} This intrinsic layer selectivity introduces a new degree of freedom: the electronic wavefunction distribution can be controlled by stacking sequence without altering chemical composition or twist angle. Such stacking-dependent wavefunction localization suggests that interfacial coupling to an adjacent semiconductor layer may depend sensitively on which atomic plane forms the interface. Therefore, TMD bilayers offer a promising route to engineer interlayer charge transfer and recombination dynamics through deterministic stacking orientation.

Here we demonstrate that stacking polarity in monolayer/bilayer TMD heterostructures enables deterministic control of interlayer photocarrier dynamics. By comparing 2H and 3R MoS$_2$ bilayers interfaced with monolayer MoSe$_2$, we show that both the interlayer charge transfer time and the interlayer exciton recombination lifetime depend sensitively on stacking sequence and interface termination. In particular, the asymmetric layer polarization unique to 3R stacking enables interface-selective carrier transfer and recombination. These results establish vertical stacking orientation as a global and chemically invariant control parameter for tailoring nonequilibrium carrier processes in van der Waals heterostructures.

\begin{figure}[t]
\centerline{\includegraphics[width=15cm]{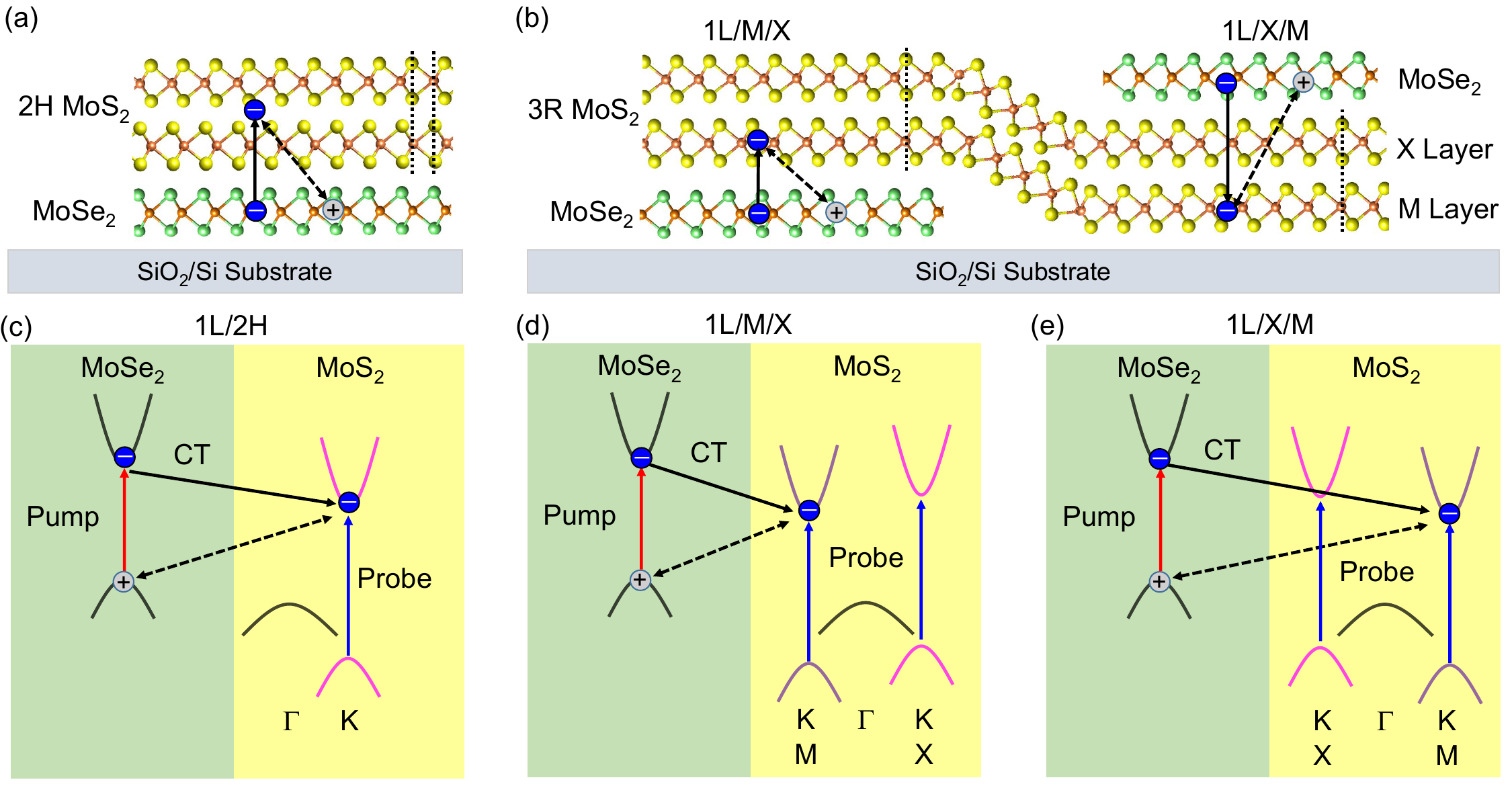}}
\caption{Illustrations of the lattice structure, band alignment, and stacking-dependent photocarrier dynamics in the investigated heterostructures. (a) Heterostructure composed of a monolayer MoSe$_2$ and a bilayer 2H MoS$_2$, denoted as 1L/2H. (b) Two heterostructures fabricated on the same 3R MoS$_2$ bilayer flake, where monolayer MoSe$_2$ interfaces with either the metal (M) layer or the chalcogen (X) layer of MoS$_2$. These configurations are labeled as 1L/M/X and 1L/X/M, respectively, highlighting the distinct interfacial MoS$_2$ layers. (c)–(e) Schematic band alignment and expected photocarrier dynamics in these heterostructures following optical excitation of the MoSe$_2$ layer, including interlayer charge transfer (CT, solid arrows) and electron (e)–hole (+) recombination (dashed arrows).} 
\label{Fig:scheme}
\end{figure}

Figure~\ref{Fig:scheme} illustrates the heterostructures investigated to probe stacking-order-dependent photocarrier dynamics in systems composed of monolayer (1L) MoSe$_2$ and bilayer (2L) MoS$_2$. Panel~(a) shows a representative heterostructure formed by a MoSe$_2$ monolayer and a 2H MoS$_2$ bilayer, hereafter denoted as 1L/2H. The corresponding schematic band alignment is shown in panel~(c). In monolayer MoSe$_2$, both the conduction band minimum (CBM) and valence band maximum (VBM) reside at the K point. For 2H MoS$_2$ bilayers, the CBM remains at the K valley and lies energetically below that of MoSe$_2$, while the VBM shifts to the $\Gamma$ point and remains lower than that of MoSe$_2$. This type-II band alignment favors electron transfer from photoexcited MoSe$_2$ into MoS$_2$, while holes remain confined in MoSe$_2$.

In the 2H bilayer, the two MoS$_2$ layers are electronically equivalent due to inversion symmetry. The two layers are rotated by $\pi$ such that the Mo (S) atoms in one layer are aligned with the S (Mo) atoms in the other layer, as indicated by the dotted lines. As a result of this symmetry, the electron wavefunction is distributed symmetrically across the two layers. Following interlayer charge transfer (CT), the transferred electron in MoS$_2$ forms an interlayer exciton with the hole remaining in MoSe$_2$, and recombination of this spatially indirect exciton determines the carrier lifetime. We fabricated devices with either MoS$_2$ placed on top of MoSe$_2$ or vice versa. Owing to the inversion symmetry of 2H MoS$_2$, these two configurations are structurally equivalent apart from their orientation relative to the SiO$_2$/Si substrate and are therefore expected to exhibit identical interlayer carrier dynamics.

In contrast, the situation differs qualitatively for heterostructures incorporating 3R-stacked MoS$_2$ bilayers, as illustrated in Figure~\ref{Fig:scheme}(b). In 3R MoS$_2$, the two layers are not rotated, and the S atoms in one layer (denoted as the X layer) align with the Mo atoms in the adjacent layer (denoted as the M layer), as indicated by the dotted lines. In contrast, the Mo atoms in that layer do not align with atoms in the neighboring layer. As a result, the M and X layers are nonequivalent, breaking inversion symmetry and generating an intrinsic out-of-plane polarization. This stacking-induced asymmetry lifts the layer degeneracy of the electronic states at the K valley: the conduction band splits into two branches predominantly localized in the M and X layers, respectively, with the M-layer branch lower in energy, as schematically shown in panels~(d) and (e).\cite{nc112391,b89075409,nn9611,np16469,sa8eade3759,nn1836} Although a similar splitting occurs in the valence band, its influence is less significant here because the VBM of bilayer MoS$_2$ lies at the $\Gamma$ point.

When a 3R MoS$_2$ bilayer forms a heterostructure with monolayer MoSe$_2$, the interface can occur at either the M layer or the X layer, as indicated in panel~(b). Because the two surfaces of a 3R bilayer cannot be optically distinguished prior to stacking, we employed a single large MoS$_2$ bilayer flake to construct both configurations. The MoS$_2$ bilayer was first transferred onto a MoSe$_2$ monolayer to form one heterostructure region. Subsequently, a second MoSe$_2$ monolayer was transferred onto a different region of the same MoS$_2$ flake. This fabrication strategy ensures that the two resulting heterostructures, labeled 1L/M/X and 1L/X/M, possess opposite stacking polarities while sharing identical MoS$_2$ thickness and crystal quality.

Panels~(d) and (e) schematically illustrate the hypothesized stacking-dependent carrier dynamics. In the 1L/M/X configuration, the lower-energy conduction-band branch localized in the M layer directly interfaces with MoSe$_2$. Electron transfer from MoSe$_2$ into MoS$_2$ therefore involves minimal spatial separation, leading to stronger interlayer coupling and faster CT. The close proximity between electrons in the interfacial M layer and holes in MoSe$_2$ is also expected to enhance interlayer recombination, shortening the carrier lifetime. In contrast, for the 1L/X/M configuration, the M layer hosting the CBM is positioned farther from the interface. The increased spatial separation reduces interlayer wavefunction overlap, which is expected to slow down the CT process and prolong the interlayer recombination lifetime. Both 3R-based heterostructures are therefore anticipated to exhibit carrier dynamics distinct from the 1L/2H case. Together, this set of heterostructures provides a controlled platform to systematically investigate stacking-order-tuned photocarrier dynamics in TMD systems.

\begin{figure}[t]
\centerline{\includegraphics[width=8.5cm]{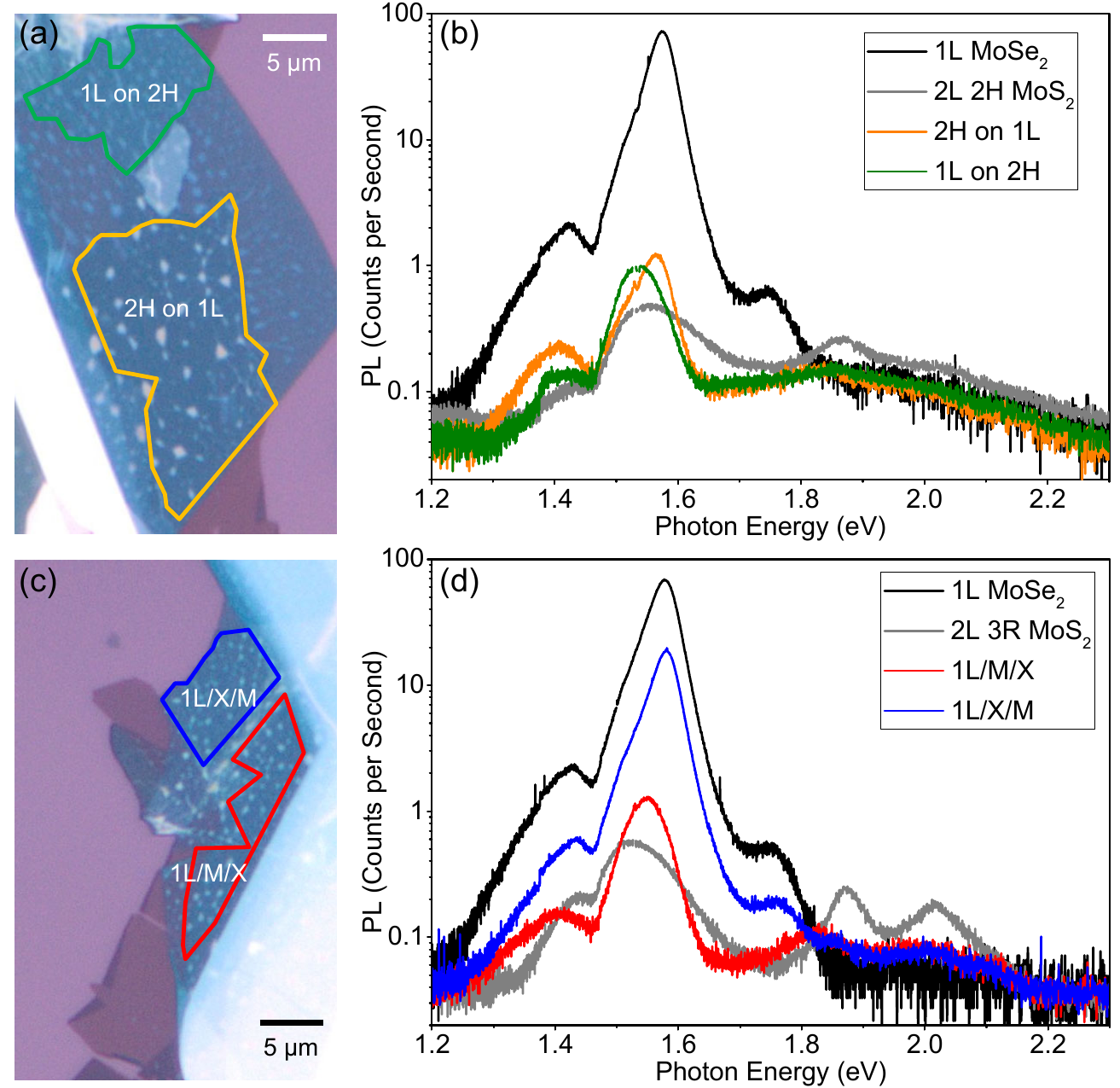}}
\caption{Sample characterization by optical microscopy and photoluminescence (PL) spectroscopy. (a) Optical microscopy image of the heterostructure composed of a monolayer (1L) MoSe$_2$ and a bilayer (2L) 2H MoS$_2$, containing regions of 1L-on-2H and 2H-on-1L stacking configurations. (b) Room-temperature PL spectra measured from different regions of the sample shown in (a). (c) Optical microscopy image of the heterostructure formed with a 3R MoS$_2$ bilayer. (d) Corresponding room-temperature PL spectra measured from selected regions of the 3R-based heterostructure.} 
\label{Fig:PL}
\end{figure}

The heterostructure samples were fabricated using standard mechanical exfoliation and dry-transfer techniques (see Supporting Information). Figure~\ref{Fig:PL}(a) shows an optical microscopy image of the heterostructures based on a 2H MoS$_2$ bilayer, 1L/2H, containing regions with two equivalent stacking configurations: 2H-on-1L and 1L-on-2H. Room-temperature photoluminescence (PL) spectroscopy was performed to characterize the samples. Under continuous-wave excitation at 3.06~eV with a power of 1~$\upmu$W, a strong PL peak centered at approximately 1.59~eV is observed from the MoSe$_2$ monolayer, whereas the 2H MoS$_2$ bilayer exhibits weaker peaks due to its indirect band structure. In the overlapping heterostructure regions, the MoSe$_2$ PL intensity is quenched by approximately two orders of magnitude, indicating efficient interlayer charge transfer which is consistent with high interfacial quality. Similarly, Figure~\ref{Fig:PL}(c) shows an optical image of the heterostructures incorporating a 3R MoS$_2$ bilayer with different interfaces, 1L/M/X and 1L/X/M. Pronounced PL quenching is also observed in the corresponding overlapping regions, as shown in Figure~\ref{Fig:PL}(d), confirming strong interlayer coupling in the 3R-based heterostructures.

The interlayer CT process in these heterostructures was investigated using the pump--probe configuration illustrated in Figure~\ref{Fig:scheme}. A 1.59~eV pump pulse selectively excites intralayer excitons in MoSe$_2$, as it is resonant with its optical bandgap. The photoexcited electrons are expected to transfer to MoS$_2$, and their dynamics are time-resolved using a 1.82~eV probe pulse tuned to the direct optical transition of MoS$_2$ at the K valley. The carrier population in MoS$_2$ is monitored via differential reflectance of the probe, defined as $\Delta R/R_0 = (R - R_0)/R_0$, where $R$ and $R_0$ denote the reflectance of the sample with and without the pump, respectively (see Supporting Information).\cite{afm271604509} Because MoS$_2$ is not directly photoexcited at 1.59~eV, the signal can be attributed to electrons transferred from MoSe$_2$. Although the K-valley transitions of the M and X layers are degenerate, and the probe therefore senses carrier populations in both layers, the signal can be primarily associated with electrons in the M layer. This is because the M layer hosts the lower-energy conduction-band state, such that electrons preferentially relax into and accumulate in this layer once quasi-equilibrium is established.

\begin{figure}[t]
\centerline{\includegraphics[width=8.5cm]{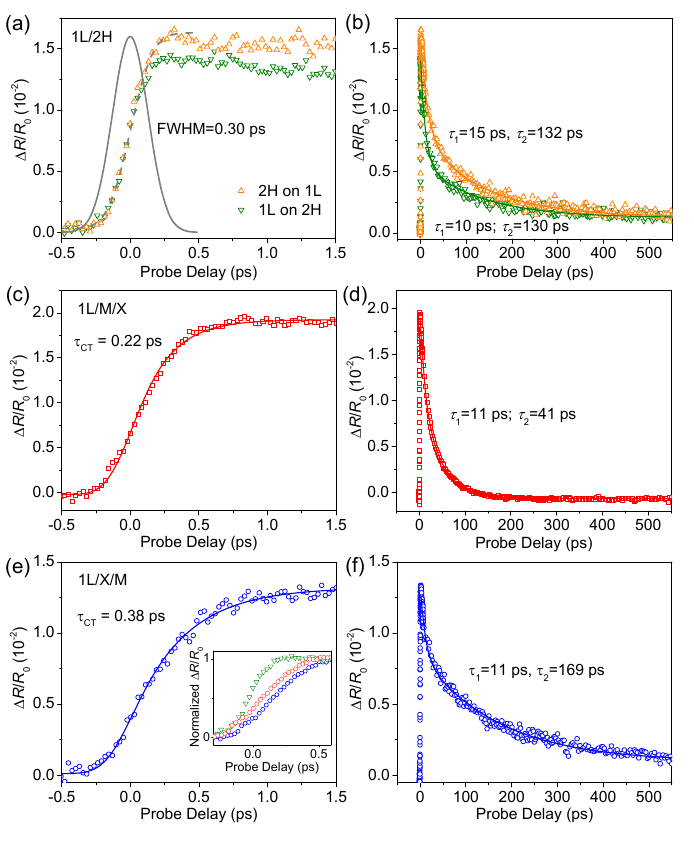}}
\caption{Photocarrier dynamics measured in heterostructures composed of monolayer MoSe$_2$ and bilayer MoS$_2$. A 1.59~eV pump selectively excites the MoSe$_2$ layer, while a 1.82~eV probe monitors the transient response of MoS$_2$. Panels (a), (c), and (e) show results from the 1L/2H, 1L/M/X, and 1L/X/M heterostructures, respectively, over a short probe-delay range. Panels (b), (d), and (f) show the corresponding datasets over a longer delay-time range. The inset in (e) provides a direct comparison of the normalized rising edges of the signals.} 
\label{Fig:CT}
\end{figure}

Figure~\ref{Fig:CT} summarizes the time-resolved differential reflectance measured from the three heterostructures at room temperature and under identical experimental conditions. In the 1L/2H heterostructure [orange symbols in Figure~\ref{Fig:CT}(a)], the rising edge of the signal is well described by the integral of a Gaussian function with a full width at half maximum (FWHM) of 0.30~ps (dashed line), corresponding to the convolution of the 0.20~ps pump and probe pulses (the solid line). This indicates that the interlayer CT occurs on a time scale shorter than the experimental temporal resolution. Such ultrafast CT is commonly observed in monolayer/monolayer TMD heterostructures with type-II band alignment.\cite{nn9682,acsnano812717,nl173591,acsnano132341,nl214738,nl155033,acsnano1112020,2dm4015033}

To exclude possible substrate effects on carrier dynamics,\cite{l105136805,apl99102109,ami61901307} we also measured a 1L/2H heterostructure with the stacking sequence inverted. As shown by the green symbols in Figure~\ref{Fig:CT}(a,b), the dynamics are indistinguishable from those of the original configuration. Because the two layers of the inversion-symmetric 2H MoS$_2$ bilayer are electronically equivalent, identical dynamics are expected. This agreement confirms that the observed carrier behavior is intrinsic to the heterostructure and not significantly influenced by the substrate.

We next examined heterostructures incorporating 3R-stacked MoS$_2$ bilayers. In contrast to the 1L/2H case, the interlayer CT becomes clearly time-resolvable. For the 1L/M/X heterostructure, the transient signal exhibits a rising edge slower than the instrumental response [Figure~\ref{Fig:CT}(c)]. Fitting the rise with a CT model obtained by convolving a single-exponential response, $\Delta R/R_0 \propto 1 - \exp[-(t - t_0)/\tau_{\mathrm{CT}}]$, with a Gaussian instrument response function (FWHM = 0.30 ps), as indicated by the solid line in Figure~\ref{Fig:CT}(c), yields $\tau_{\mathrm{CT}} = 0.22$~ps. In the 1L/X/M heterostructure, the CT time further increases to $\tau_{\mathrm{CT}} = 0.38$~ps [Figure~\ref{Fig:CT}(e)].

Because CT in the 1L/2H heterostructure is faster than the temporal resolution, these results demonstrate that stacking order tunes the CT time by several-fold. Such variation can influence CT efficiency, as interlayer CT competes with ultrafast carrier cooling and intralayer recombination. Moreover, stacking-controlled CT directly impacts the temporal response of optoelectronic devices. 

The stacking dependence of CT originates from symmetry-controlled wavefunction localization. In the inversion-symmetric 2H bilayer, the electron wavefunction at the K valley extends across both layers,\cite{b89075409} resulting in strong interfacial electronic coupling and ultrafast CT, similar to monolayer/monolayer heterostructures.\cite{nn9682,acsnano812717,nl173591,acsnano132341,nl214738,nl155033,acsnano1112020,2dm4015033} In contrast, broken inversion symmetry in 3R MoS$_2$ lifts the layer degeneracy and produces layer-polarized conduction-band states, in which the two branches are predominantly localized in different layers, with the lower-energy branch residing primarily in the M layer.\cite{b89075409,nn9611,x12041005} The energy splitting between these branches is on the order of tens of meV,\cite{b89075409,x12041005} which can significantly influence exciton dynamics at room temperature.\cite{b109035410} When forming heterostructures, the layer-selective electronic structure in 3R MoS$_2$ reduces the effective interfacial coupling with MoSe$_2$, thereby slowing CT. The further reduction of the CT rate in 1L/X/M compared to 1L/M/X arises from the increased spatial separation between the electron-localized M layer and the interface, as illustrated in Figure~\ref{Fig:scheme}(d,e), due to the conduction-band splitting.

The stacking-dependent CT time observed here represents a rare case in which interlayer CT becomes directly time-resolvable, providing a new platform to investigate CT dynamics in TMD heterostructures. In general, factors such as strain, interfacial disorder, and twist-angle variations are expected to influence the CT time. However, in previously studied monolayer/monolayer heterostructures, CT is ultrafast and often occurs faster than the experimental temporal resolution, remaining effectively unresolvable across a wide range of twist angles, temperatures, dielectric environments, and excitation conditions.\cite{acsnano106612,acsnano1112020,nl173591,jpcl10150,sa5eaau0073} In contrast, the systematic and reproducible differences observed here demonstrate that stacking polarity provides a unique and effective means to tune and directly resolve interlayer CT processes. This capability could enable further investigations of how these factors influence the CT time.

The decay dynamics, shown in the right column of Figure~\ref{Fig:CT}, reveal pronounced stacking-dependent tuning of the interlayer exciton lifetime. The decay traces are well fitted by a biexponential function. For the two 1L/2H heterostructure, the extracted time constants are comparable. The longer component $\tau_2$ of about 130 ps is attributed to recombination of interlayer excitons formed by transferred electrons in MoS$_2$ and holes in MoSe$_2$. In the 1L/M/X heterostructure, the interlayer exciton lifetime decreases to 41~ps, consistent with enhanced electron–hole wavefunction overlap at the interface. In contrast, the 1L/X/M heterostructure exhibits a prolonged lifetime of 169~ps, corresponding to increased electron–hole separation. These results demonstrate that stacking order tunes the interlayer exciton lifetime by several-fold, paralleling the systematic control observed for the CT dynamics. 

An additional decay channel ($\tau_1$) in the range of 10–15 ps is observed in all heterostructures. This component is attributed to intralayer dynamics in MoSe$_2$, arising from regions with reduced interlayer coupling. In TMD heterostructures fabricated by exfoliation and transfer techniques, similar dynamics have been reported and are commonly associated with areas of imperfect interfacial contact. \cite{acsnano812717,nl254054,apl118253106} In such regions, photoexcited carriers do not efficiently undergo interlayer charge transfer and instead relax as in isolated monolayers. These regions are typically attributed to trapped residues or interfacial contamination introduced during fabrication. Thermal annealing can cause such residues to migrate and aggregate, leading to the formation of regions with strong interlayer coupling as well as regions with reduced coupling that behave more like isolated monolayers. \cite{nc95387,acsnano1314182} The similar time constants observed across all heterostructures further support this assignment.

\begin{figure}[t]
\centerline{\includegraphics[width=15cm]{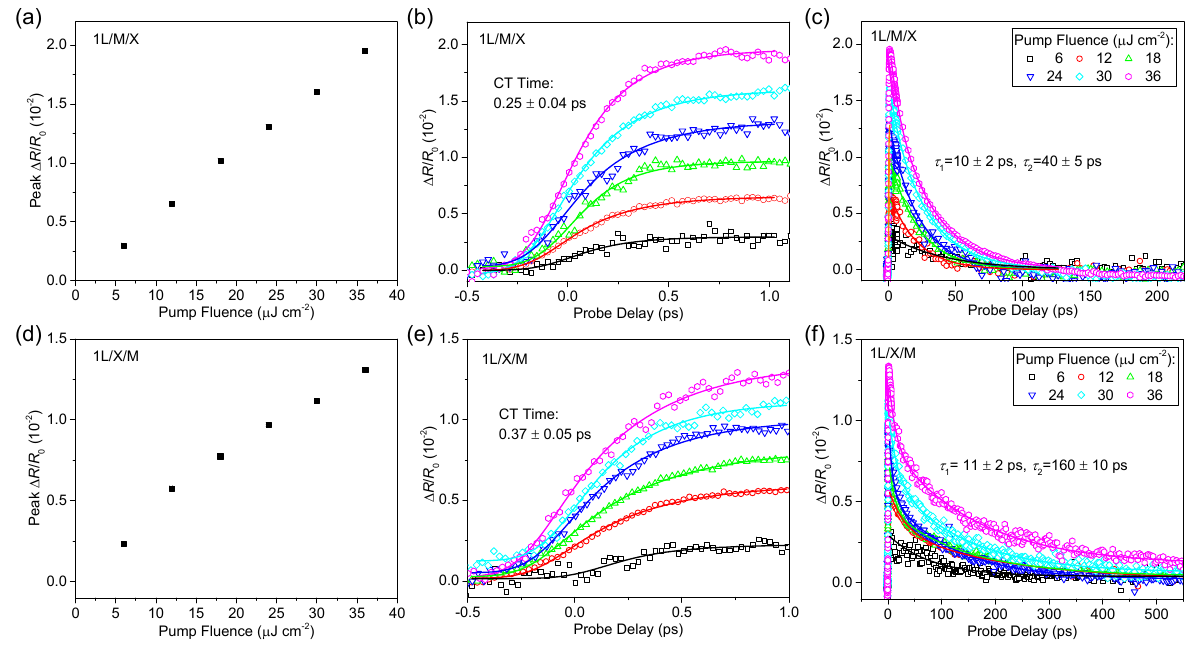}}
\caption{Pump-fluence dependence of the photocarrier dynamics measured in heterostructures incorporating 3R MoS$_2$ bilayers using a 1.59~eV pump and a 1.82~eV probe. (a) Peak differential reflectance near zero probe delay as a function of pump fluence for the 1L/M/X heterostructure, demonstrating a linear fluence dependence. (b) and (c) Time-resolved differential reflectance with various pump fluence and fits (see text), showing minimal dependence of the dynamics on pump fluence. (d)–(f) Corresponding measurements for the 1L/X/M heterostructure.} 
\label{Fig:CTpower}
\end{figure}

To validate our interpretation of the time-resolvable CT process in 3R MoS$_2$–based heterostructures and to exclude possible effects arising from different injected carrier densities associated with stacking order, we performed measurements on the 1L/M/X and 1L/X/M samples over a range of pump fluences. The results are summarized in Figure~\ref{Fig:CTpower}. The peak differential reflectance near zero delay scales linearly with pump fluence for both heterostructures [Figure~\ref{Fig:CTpower}(a,d)], confirming that the signal is proportional to the carrier density. The rising edge of the signal, modeled using the CT model convolved with the instrument response function, remains unchanged with fluence. The extracted CT times [Figure~\ref{Fig:CTpower}(b,e)] show no systematic fluence dependence, yielding average values of $0.25 \pm 0.04$~ps and $0.37 \pm 0.05$~ps for the 1L/M/X and 1L/X/M heterostructures, respectively. Likewise, the interlayer exciton recombination lifetimes exhibit no observable dependence on pump fluence, as shown in Figure~\ref{Fig:CTpower}(c,f). These results confirm that the stacking-dependent CT and recombination dynamics arise from structural differences rather than carrier-density effects.

In summary, we demonstrate that vertical stacking order provides a deterministic and chemically invariant means to control interlayer photocarrier dynamics in TMD heterostructures. By comparing 2H and 3R MoS$_2$ bilayers interfaced with monolayer MoSe$_2$, we show that both the interlayer charge transfer time and the interlayer exciton recombination lifetime can be tuned by several-fold through stacking polarity. The broken inversion symmetry and resulting layer-polarized electronic states in 3R MoS$_2$ enable interface-selective interlayer coupling, enabling systematic control of electron transfer and recombination without modifying chemical composition or introducing twist-angle–induced spatial inhomogeneity. These findings establish stacking order as a global control parameter for nonequilibrium carrier dynamics and provide a practical pathway for engineering ultrafast charge separation, carrier lifetime, and device response time in van der Waals optoelectronic systems.

\begin{acknowledgement}

Research supported by the U.S. Department of Energy, Office of Basic Energy Sciences, Division of Materials Sciences and Engineering under Award DE-SC0020995.

\end{acknowledgement}

\begin{suppinfo}

Additional experimental details, including sample fabrication, photoluminescence, and ultrafast pump--probe techniques. 

\end{suppinfo}


\providecommand{\latin}[1]{#1}
\makeatletter
\providecommand{\doi}
  {\begingroup\let\do\@makeother\dospecials
  \catcode`\{=1 \catcode`\}=2 \doi@aux}
\providecommand{\doi@aux}[1]{\endgroup\texttt{#1}}
\makeatother
\providecommand*\mcitethebibliography{\thebibliography}
\csname @ifundefined\endcsname{endmcitethebibliography}  {\let\endmcitethebibliography\endthebibliography}{}

\end{document}